\documentclass[prd,superscriptaddress,unsortedaddress,twocolumn,preprintnumbers,amsmath,amssymb]{revtex4}

\usepackage[dvipdfmx]{graphicx}
\usepackage{amsmath,amssymb,times}
\usepackage{color}
\usepackage{ulem}
\usepackage{bm}

\def\fun#1#2{\lower3.6pt\vbox{\baselineskip0pt\lineskip.9pt
\ialign{$\mathsurround=0pt#1\hfil##\hfil$\crcr#2\crcr\sim\crcr}}}

\newcommand{\beq}{\begin{equation}}
\newcommand{\eeq}{\end{equation}}
\newcommand{\bea}{\begin{eqnarray}}
\newcommand{\eea}{\end{eqnarray}}

\newcommand{\la}{\left\langle}

\def\gsim{~\,\makebox(1,1){$\stackrel{>}{\widetilde{}}$}\,~}
\def\lsim{~\,\makebox(1,1){$\stackrel{<}{\widetilde{}}$}\,~}

\DeclareSymbolFont{boldletters}{OML}{cmm} {b}{it}
\DeclareSymbolFontAlphabet{\mathbit}{boldletters}
\DeclareMathSymbol{\alpha}{\mathalpha}{letters}{"0B}
\DeclareMathSymbol{\beta}{\mathalpha}{letters}{"0C}
\DeclareMathSymbol{\gamma}{\mathalpha}{letters}{"0D}
\DeclareMathSymbol{\delta}{\mathalpha}{letters}{"0E}
\DeclareMathSymbol{\epsilon}{\mathalpha}{letters}{"0F}
\DeclareMathSymbol{\zeta}{\mathalpha}{letters}{"10}
\DeclareMathSymbol{\eta}{\mathalpha}{letters}{"11}
\DeclareMathSymbol{\theta}{\mathalpha}{letters}{"12}
\DeclareMathSymbol{\iota}{\mathalpha}{letters}{"13}
\DeclareMathSymbol{\kappa}{\mathalpha}{letters}{"14}
\DeclareMathSymbol{\lambda}{\mathalpha}{letters}{"15}
\DeclareMathSymbol{\mu}{\mathalpha}{letters}{"16}
\DeclareMathSymbol{\nu}{\mathalpha}{letters}{"17}
\DeclareMathSymbol{\xi}{\mathalpha}{letters}{"18}
\DeclareMathSymbol{\pi}{\mathalpha}{letters}{"19}
\DeclareMathSymbol{\rho}{\mathalpha}{letters}{"1A}
\DeclareMathSymbol{\sigma}{\mathalpha}{letters}{"1B}
\DeclareMathSymbol{\tau}{\mathalpha}{letters}{"1C}
\DeclareMathSymbol{\upsilon}{\mathalpha}{letters}{"1D}
\DeclareMathSymbol{\phi}{\mathalpha}{letters}{"1E}
\DeclareMathSymbol{\chi}{\mathalpha}{letters}{"1F}
\DeclareMathSymbol{\psi}{\mathalpha}{letters}{"20}
\DeclareMathSymbol{\omega}{\mathalpha}{letters}{"21}
\DeclareMathSymbol{\varepsilon}{\mathalpha}{letters}{"22}
\DeclareMathSymbol{\vartheta}{\mathalpha}{letters}{"23}
\DeclareMathSymbol{\varpi}{\mathalpha}{letters}{"24}
\DeclareMathSymbol{\varrho}{\mathalpha}{letters}{"25}
\DeclareMathSymbol{\varsigma}{\mathalpha}{letters}{"26}
\DeclareMathSymbol{\varphi}{\mathalpha}{letters}{"27}
\DeclareMathSymbol{\Gamma}{\mathalpha}{letters}{"00}
\DeclareMathSymbol{\Delta}{\mathalpha}{letters}{"01}
\DeclareMathSymbol{\Theta}{\mathalpha}{letters}{"02}
\DeclareMathSymbol{\Lambda}{\mathalpha}{letters}{"03}
\DeclareMathSymbol{\Xi}{\mathalpha}{letters}{"04}
\DeclareMathSymbol{\Pi}{\mathalpha}{letters}{"05}
\DeclareMathSymbol{\Sigma}{\mathalpha}{letters}{"06}
\DeclareMathSymbol{\Upsilon}{\mathalpha}{letters}{"07}
\DeclareMathSymbol{\Phi}{\mathalpha}{letters}{"08}
\DeclareMathSymbol{\Psi}{\mathalpha}{letters}{"09}
\DeclareMathSymbol{\Omega}{\mathalpha}{letters}{"0A}

\def\la{\mathrel{\mathpalette\fun <}}

\def\fun#1#2{\lower3.6pt\vbox{\baselineskip0pt\lineskip.9pt
\ialign{$\mathsurround=0pt#1\hfil##\hfil$\crcr#2\crcr\sim\crcr}}}

\begin{document}
\title{Extrapolation for meson screening masses \\ 
from imaginary to real chemical potential}

\author{Masahiro Ishii}
\email[]{ishii@phys.kyushu-u.ac.jp}
\affiliation{Department of Physics, Graduate School of Sciences, Kyushu University,
             Fukuoka 819-0395, Japan}             

\author{Akihisa Miyahara}
\email[]{miyahara@email.phys.kyushu-u.ac.jp}
\affiliation{Department of Physics, Graduate School of Sciences, Kyushu University,
             Fukuoka 819-0395, Japan}

\author{Hiroaki Kouno}
\email[]{kounoh@cc.saga-u.ac.jp}
\affiliation{Department of Physics, Saga University,
             Saga 840-8502, Japan}  

\author{Masanobu Yahiro}
\email[]{yahiro@phys.kyushu-u.ac.jp}
\affiliation{Department of Physics, Graduate School of Sciences, Kyushu University,
             Fukuoka 819-0395, Japan}             

\date{\today}

\begin{abstract}
We first extend our formulation for the calculation of $\pi$- and
 $\sigma$-meson screening masses to the case
 of finite chemical potential $\mu$. We then consider the
 imaginary-$\mu$ approach, which is an extrapolation method from
 imaginary chemical potential ($\mu=i \mu_{\rm I}$) to real one ($\mu=\mu_{\rm R}$). The feasibility of the
 method is discussed based on the entanglement
 Polyakov-loop extended Nambu--Jona-Lasinio (EPNJL) model in 2-flavor
 system. As an example, we investigate 
how reliable the imaginary-$\mu$ approach is 
for $\pi$- and $\sigma$-meson screening masses, 
comparing ``screening  masses at $\mu_{\rm R}$ in the method'' with
 ``those calculated directly at $\mu_{\rm R}$''. We finally propose the new extrapolation method and confirm its efficiency. 
\end{abstract}

\maketitle

\section{Introduction}
\label{Introduction}

$T$ and $\mu$ dependence of hadron masses are closely related 
with those of the ground-state structure of hot QCD matter, 
where $T$ is temperature and $\mu$ means quark-number chemical
potential. 
In fact, medium modification of vector and $\eta'$ mesons has been measured in 
heavy-ion collisions~\cite{Csorgo,STAR:2014}. These results indicate 
the chiral and the effective $U(1)_{\rm A}$-symmetry restoration. 
It is, therefore,  important to determine $T$ and $\mu$  dependence of 
light hadron masses.

Lattice QCD (LQCD) is powerful tool to investigate the QCD matter
at finite $T$ and $\mu$. In fact, many LQCD calculations have been done for low
density ($\mu/T\lsim 1$). The calculation in high density region is still challenging because of well-known ``sign problem''.
Several methods were proposed so far to circumvent the sign problem;  
the Taylor expansion method~\cite{Allton,Ejiri},
the reweighting method~\cite{Fodor-Katz-reweighting}, 
the imaginary-$\mu$ method~\cite{Forcrand-Philipsen,DElia-Lombardo,Nagata-Nakamura,Sugano:2017yqz}, 
the canonical approach~\cite{Nakamura}, 
the complex Langevin method~\cite{Aarts1, Aarts2, Sexty, Aarts3}, 
and the Lefschetz thimble theory~\cite{Cristoforetti, Fujii}. 
These have made great progress, but all the results are consistent 
only in $\mu/T \la 1$ at the present stage. 
Among them,  we pick up the the imaginary-$\mu$ method in the present
paper. When one considers $\mu$ as complex variable, this method corresponds to the analytic continuation from 
the imaginary chemical potential ($\mu=i \mu_{\rm I}$) to the real one 
($\mu=\mu_{\rm R}$).

In LQCD simulation for finite $\theta \equiv \mu_{\rm I}/T$, 
the thermodynamic potential $\Omega_{\rm QCD}(\theta)$ 
has the Roberge and Weiss (RW) periodicity: $\Omega_{\rm QCD}(\theta)=\Omega_{\rm QCD}(\theta+2\pi/3)$~\cite{RW}.  
The QCD phase diagram has 
the first-order phase transition (RW phase transition) at $T \ge T_{\rm RW}$ and 
$\theta= \pi/3$, where $T_{\rm RW}$ is RW transition temperature. The endpoint of RW phase transition is located at 
$(\theta,T)=(\pi /3,T_{\rm RW})$. The order of the RW endpoint and the value of $T_{\rm RW}$ have been
investigated in 2-flavor LQCD
simulations~\cite{Forcrand-Philipsen,DElia-Lombardo,Nagata-Nakamura}.

One can consider effective models as an complementary approach to 
the first-principle LQCD simulation. 
The Polyakov-loop extended Nambu--Jona-Lasinio (PNJL) model~\cite{Meisinger,Dumitru,Fukushima1,Costa:2005,Ghos,Megias,Ratti1,Ciminale,Ratti2,Rossner,Hansen,Sasaki-C,Schaefer,Kashiwa1,Sakai1,Sakai2,Kouno,Sakai:2009dv,Sakai_JPhys,Costa:2009,Sakai:2010kx}
qualitatively reproduces 2-flavor LQCD data 
in $\mu_{\rm R}/T \la 1$, since the model can treat  
the chiral and the deconfinement transition simultaneously. 
In addition, the  model is successful in accounting for 2-flavor LQCD data 
in $0\le \theta \la \pi/3$~\cite{Kouno,Sakai:2009dv}, because it has 
the RW periodicity. The entanglement PNJL (EPNJL)
model~\cite{Sakai_EPNJL,Sakai:2011fa,Ruivo:2012a} is improved version of PNJL model.
The EPNJL
model quantitatively
reproduces 2-flavor LQCD data  in $0\le \theta \la \pi/3$~\cite{Sakai_EPNJL} and $\mu_{\rm R}/T \la 1$~\cite{Sakai:2011fa}, 
since the model possesses the RW periodicity and the strong correlation
between the chiral and the deconfinement
transition. 

Meson masses can be classified into ``meson pole mass'' and 
``meson screening mass''. 
In LQCD simulations at finite $T$, 
the derivation of meson screening mass is easier than that of 
meson pole mass, 
since the spatial lattice size is larger than  the temporal one;
see Appendix of Ref.~\cite{Ishii:2016dln} for the further explanation. 
Meanwhile, in NJL-type effective models, time-consuming 
calculations were needed for the meson screening mass compared with that
of the meson pole mass. 
Recently, this problem was solved  
by our previous works~\cite{Ishii:2013kaa,Ishii:2015ira,Ishii:2016dln} for the case of $\mu=0$.

In this paper, for simplicity, we concentrate on the $\pi$-meson and $\sigma$-meson screening masses, $M_{\pi}^{\rm scr}$ and $M_{\sigma}^{\rm scr}$, 
in the framework of 2-flavor EPNJL model. 
We  apply the method of Ref.~\cite{Ishii:2013kaa,Ishii:2015ira,Ishii:2016dln} 
for the case of finite $\mu_{\rm R}$ and $\mu_{\rm I}$, 
and then investigate how reliable the imaginary-$\mu$ method is 
for $M_{\pi}^{\rm scr}$ and $M_{\sigma}^{\rm scr}$. 
For this purpose, we compare ``
the $M_{\xi}^{\rm scr}$ extrapolated from  $i \mu_{\rm I}$ ({\bf extrapolating result})'' with 
``the $M_{\xi}^{\rm scr}$ calculated directly at $\mu_{\rm R}$ ({\bf direct result})'' for $\xi=\pi,\sigma$ mesons.

In Sec.~\ref{Formalism}, we explain a way of calculating the 
meson screening mass at finite $\mu$. 
Numerical results are shown in Sec. \ref{Numerical Results}. 
Section \ref{Summary} is devoted to a summary. 

\newpage

\section{Formalism}
\label{Formalism}

\subsection{Model setting}
\label{Model-setting}
The Lagrangian density of 2-flavor EPNJL model is defined by
\begin{align}
 {\cal L}  
=& {\bar \psi}(i \gamma_\nu D^\nu -m_0)\psi  + G_{\rm S}(\Phi)[({\bar \psi}\psi )^2 
  +({\bar \psi }i\gamma_5 {\vec \tau}\psi )^2]
\nonumber\\
 & -{\cal U}(\Phi [A],{\bar \Phi} [A],T) 
\label{L}
\end{align}
with u- and d-quark fields $\psi = (u,d)^{\rm T}$ and the isospin matrix
${\vec \tau}$. We assume isospin symmetry, i.e., u and d quarks have the same mass $m_0$. The gluon field $A^\nu$ is introduced
through the covariant derivative $D^\nu=\partial^\nu + iA^\nu$ with $A^\nu=\delta^{\nu}_{0}g(A^0)_a{\lambda_a/2}
=-\delta^{\nu}_{0}ig(A_{4})_a{\lambda_a/2}$, where the matrices $\lambda_a$ are the Gell-Mann matrices in color space and $g$ is the gauge
coupling. Here, we consider only the time component $A_4$ of $A_\nu$ 
and assume that the $A_4$ is a homogeneous and static 
background field. 

In the EPNJL model, the Polyakov loop $\Phi$ and 
its Hermitian conjugate ${\bar \Phi}$ are defined by   
\begin{align}
\Phi &= {1\over{3}}{\rm tr}_{\rm c}(L),
~~~~~{\bar \Phi} ={1\over{3}}{\rm tr}_{\rm c}({L^*})
\label{Polyakov}
\end{align}
with $L= \exp[i A_4/T]=\exp[i\hspace{0.5ex}{\rm diag}(A_4^{11},A_4^{22},A_4^{33})/T]$ 
for real classical variables $A_4^{jj}$ ($j=1,2,3$). The trace ${\rm
tr}_{\rm c}$ is taken in color space. The relation
between $A_4^{jj}$ and 
$\Phi$ or $\bar{\Phi}$ is summarized in Appendix~\ref{A4-from-Polyakov-loop}.
The coupling constant $G_{\rm S}$ of the 
four-quark interaction is assumed to depend on the Polyakov loop $\Phi$
and $\bar{\Phi}$;  
\begin{equation}
G_{\rm S}(\Phi)=G_{\rm S}(0)\times\left[1-\alpha_1\Phi{\bar \Phi} -\alpha_2\left(\Phi^3 + {\bar \Phi}^{3}\right)\right].
\label{EPNJL}
\end{equation}
We set the parameters $\alpha_1,\alpha_2$ to $\alpha_1=\alpha_2=0.2$ to reproduce LQCD data on $T$ dependence of chiral 
condensate~\cite{Chiral-condensate-Karsch} and Polyakov 
loop~\cite{Polyakov-loop-Kaczmarek}; see Sec.~\ref{Chiral and
Polyakov-loop} for the further explanation. 

The Polyakov loop $\Phi$ and its Hermitian conjugate $\bar{\Phi}$
are mainly governed by the Polyakov-loop potential~$\mathcal{U}$ in Eq.~\eqref{L}. We use the logarithm-type Polyakov-loop potential
$\mathcal{U}$ of Ref.~\cite{Rossner}. The parameter set in $\mathcal{U}$
is determined from LQCD data on thermodynamic quantities in the pure
gauge limit. The $\mathcal{U}$ has one dimensionful parameter
$T_0$ and the value is often set to $T_0=270$ MeV since the deconfinement
transition occurs at $T=270$ MeV in the pure gauge limit. When one considers
the dynamical quarks, the typical energy scale $T_0$ depends on 
the number  of flavors ($N_{\rm f}$). Hence we treat $T_0$
as an adjustable parameter and determine the value to reproduce the pseudocritical temperature $T_{\rm c}^{\chi} = 173\pm
8$~MeV for chiral transition in 2-flavor LQCD simulations at zero
chemical potential~\cite{Chiral-condensate-Karsch,Polyakov-loop-Kaczmarek,Tc-Karsch}.
The parameter thus obtained is $T_0=200$~MeV.

Applying the mean field approximation to Eq. \eqref{L} leads to 
the linearized Lagrangian density 
\begin{align}
 {\cal L}^{\rm MFA}  
= {\bar \psi}S^{-1}\psi  - G_{\rm S}(\Phi)\sigma^2  - 
{\cal U}(\Phi [A],{\bar \Phi} [A],T), 
\label{linear-L}
\end{align} 
where the dressed quark propagator $S$ is defined by 
\bea
S=\frac{1}{i \gamma_\nu \partial^\nu - i\gamma_0A_4 -M} 
\label{propa}
\eea
with the effective quark mass $M=m_0-2G_{\rm S}(\Phi)\sigma$ and 
the chiral condensate $\sigma=\langle\bar{\psi}\psi\rangle$. 
One can make the path integral over the quark fields analytically, and the thermodynamic
potential $\Omega$ per unit volume is obtained by 
\begin{align}
&\Omega\nonumber
= U_{\rm M}+{\cal U}-2 N_{\rm f} \int \frac{d^3 {\bm p}}{(2\pi)^3}
   \Bigl[ 3 E_{\bm p} \notag \\
&+ \frac{1}{\beta}
           \ln~ [1 + 3(\Phi+{\bar \Phi} e^{-\beta (E_{\bm p}-\mu )}) 
           e^{-\beta (E_{\bm p}-\mu)}+ e^{-3\beta (E_{\bm p}-\mu)}] \notag\\
&+ \frac{1}{\beta} 
           \ln~ [1 + 3({\bar \Phi}+{\Phi e^{-\beta (E_{\bm p}+\mu)}}) 
              e^{-\beta (E_{\bm p}+\mu)}+ e^{-3\beta (E_{\bm p}+\mu)}]
	      \Bigl]
	      \nonumber\\
\label{EPNJL-Omega}
\end{align}
with $E_{\bm p}=\sqrt{{\bm p}^2+M^2}$ and 
$U_{\rm M}=G_{\rm S}(\Phi)\sigma^2$. 
The mean-field variables $\sigma,\Phi,\bar{\Phi}$ are
determined so as to minimize the potential $\Omega$. 
For real $\mu$, we take the approximation 
$\Phi=\bar{\Phi}$ for simplicity.  
This approximation is pretty good for $\mu_{\rm R}/T\lsim 1$ and not so bad 
even for $\mu_{\rm R}/T\gsim 1$~\cite{Sakai:2010kx}.

In the $\mu_{\rm I}$ region, this thermodynamic potential $\Omega$ has the RW periodicity~\cite{Sakai1,Sakai2,Sakai_JPhys}. The RW
periodicity stems from the fact that $\Omega$ is
invariant under the extended $Z_3$ transformation~\cite{Sakai_JPhys} defined 
by 
\begin{eqnarray}
 \Phi \to e^{-i2\pi k/3}\Phi,~\bar{\Phi} \to e^{i2\pi k/3}\bar{\Phi},~
\theta \to \theta + \frac{2\pi k}{3}
\label{extended-Z3}
\end{eqnarray}
for integer $k$.

The three-dimensional momentum ${\bm p}$ integral 
in Eq.~\eqref{EPNJL-Omega} has ultraviolet
divergence and needs to be regularized. In this paper, we use the 
Pauli--Villars (PV) regularization~\cite{Florkowski,PV}. 
When $\Omega$ is divided into $\Omega = U_{\rm M} + \mathcal{U} + \Omega_{\rm F}(M)$, the function $\Omega_{\rm F}(M)$ is regularized in the PV scheme as 
\begin{eqnarray}
  \Omega^{{\rm reg}}_{\rm F}(M)&=&\sum_{\alpha=0}^2 C_\alpha \Omega_{\rm F}(M_{\alpha}),
\label{PV}
\end{eqnarray}
where $M_{0}=M$ and the $M_{\alpha}~(\alpha =1, 2)$ mean masses 
of auxiliary particles. The parameters $M_{\alpha}$ and
$C_\alpha$ are determined so as to satisfy the condition  
$\sum_{\alpha=0}^2C_\alpha=\sum_{\alpha=0}^2 C_\alpha
M_{\alpha}^2=0$ in order to remove the quartic, the quadratic and 
the logarithmic divergence in $\Omega_{\rm F}$. 
We then set $(C_0,C_1,C_2)=(1,-2,1)$ and 
$(M_0^2,M_{1}^2,M_{2}^2)=(M^2,M^2+\Lambda^2,M^2+2\Lambda^2)$. 
The parameter $\Lambda$ should be finite even after the regularization \eqref{PV}, since the present model is 
non-renormalizable. 

The EPNJL model has three parameters $m_0,G_{\rm S}(0),\Lambda$ in addition
to $T_0,\alpha_1,\alpha_2$. We set $m_0$ to $m_0=6.3$ MeV and determine
$G_{\rm S}(0),\Lambda$ to reproduce the experimental values of pion mass
$M_\pi=138$ MeV and its decay constant $f_\pi=93.3$ MeV at vacuum. The
EPNJL model parameters are summarized in Table~\ref{Model parameters}.
 
\begin{table}[h]
\begin{center}
\caption
{Model parameters}

\begin{tabular}{lcccccc}
\hline\hline
$m_0$~[{\rm MeV}]
&$\Lambda$~[{\rm MeV}]
&$G_{\rm S}(0)\Lambda^2$
&$\alpha_1$
&$\alpha_2$
&$T_0$~[{\rm MeV}]
\\
\hline
6.3
&768
&2.95
&0.2
&0.2
&200
\\
\hline
\end{tabular}
 \label{Model parameters}
\end{center}
\end{table}

\subsection{Meson screening mass at finite $T$ and $\mu$}
\label{Meson-mass}

Following the previous work~\cite{Ishii:2013kaa}, we first consider $\pi$ and $\sigma$ mesons at $T=\mu=0$. The current operator is expressed by  
\beq
  J_{\xi}(x) = \bar \psi(x)\Gamma_\xi \psi(x)
             - \langle\bar \psi(x) \Gamma_\xi \psi(x)\rangle  
\label{source}
\eeq
with $x=(t,{\bm x})$ for meson species $\xi=\pi,\sigma$, where $\Gamma_\sigma = 1$ for $\sigma$ meson and
$\Gamma_\pi=i\gamma_5\tau_3$ for $\pi$ meson. 
The mesonic correlation function in
coordinate space is defined by 
\beq
\zeta_{\xi\xi} (t,{\bm x}) \equiv \langle 0 | {\rm T} \left( J_\xi(t,{\bm x})
J^{\dagger}_{\xi}(0) \right) | 0 \rangle.
\label{eq:zeta}
\eeq
Here, the symbol ${\rm T}$ stands for the time-ordered
product. The Fourier transform $\chi_{\xi\xi} (q_0^2,{\bm q}^2)$ of 
$\zeta_{\xi\xi} (t,{\bm x})$ is obtained by  
\beq
\chi_{\xi\xi} (q^2_0, {\tilde q}^2) 
=  i \int d^4x ~e^{i q\cdot x} \zeta_{\xi\xi} (t,{\bm x}) 
\eeq
for an external momentum $q=(q_0,{\bm q})$ and $\tilde{q}=\pm |{\bm
q}|$. When we take the random-phase approximation, 
we can get $\chi_{\xi\xi}$ as
\begin{equation}
\chi_{\xi\xi} = \frac{\Pi_{\xi}}{1-2G_{\rm
S}(\Phi)\Pi_{\xi}} 
\label{chixixi}
\end{equation}
for $\xi=\pi,\sigma$. 
The one-loop polarization function $\Pi_{\xi}$ is explicitly calculated by 
\begin{eqnarray}
\Pi_{\sigma}
&=&(-2i) \int \frac{d^4 p}{(2\pi)^4} 
{\rm tr}_{\rm c,d} \left(iS(p+q)iS(p)\right)
\nonumber\\
&=&4i[I_{1} + I_{2}-\left(q^2 - 4M^2\right)I_{3}] 
\label{Pi_S}
\end{eqnarray}
for $\sigma$ meson and 
\begin{eqnarray}
 \Pi_{\pi}
  &=&(-2i) \int \frac{d^4 p}{(2\pi)^4} 
{\rm tr}_{\rm c,d} \left((i\gamma_5)iS(p+q)(i\gamma_5)iS(p)\right)
\nonumber\\
  &=&4i[I_{1} + I_{2} - q^2I_{3}]
\label{Pi_P}
\end{eqnarray}
for $\pi$ meson,
where the trace ${\rm tr}_{\rm c,d}$ is taken in color and Dirac
spaces. Three functions in Eqs.~\eqref{Pi_S} and~\eqref{Pi_P} are
defined by
\bea
I_{1}&=&\int {d^4p\over{(2\pi )^4}}{\rm tr_c}\Bigl[{1\over{p^2-M^2}}\Bigr],
\label{I1}
\\
I_{2}&=&\int {d^4p\over{(2\pi )^4}}{\rm tr_c}\Bigl[{1\over{(p+q)^2-M^2}}\Bigr],\label{I2}
\\
I_{3}&=&\int {d^4p\over{(2\pi )^4}}{\rm
 tr_c}\Bigl[{1\over{(p^2-M^2)((p + q)^2-M^2)}}\Bigr] .~~~
\label{I3}
\\
\nonumber
\eea
These functions are regularized with the same procedure as shown 
in Eq.~\eqref{PV}.  

In the two cases of (a) finite $T$ and $\mu=\mu_{\rm R}$ and (b) finite $T$ and $\mu=i\mu_{\rm I}$, one can get the final equations by taking the following replacement
\begin{align}
&p_0 \to i \omega_n + iA_{4} + \mu = i(2n+1) \pi T + iA_{4} + \mu, 
\nonumber\\
&\int \frac{d^4p}{(2 \pi)^4} 
\to iT\sum_{n=-\infty}^{\infty} \int \frac{d^3{\bm p}}{(2 \pi)^3}. 
\label{finte_T_mu}
\end{align}

\begin{figure}[htbp]
\begin{center}
  \includegraphics[width=0.3\textwidth]{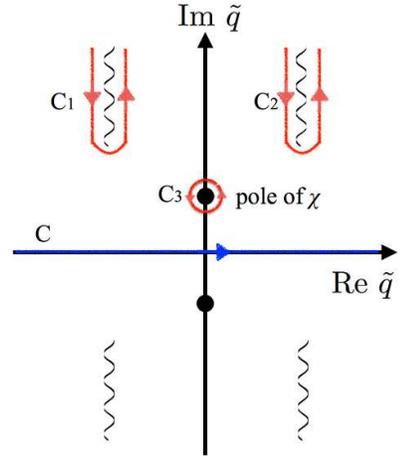}
\end{center}
\caption{Singularities of $\chi_{\xi\xi}(0,\tilde{q}^2)$ 
in the complex-$\tilde{q}$ plane. Cuts are denoted by the wavy lines and
 poles are denoted by points. The threshold masses correspond to the endpoints of cuts.
The original contour $C$ in Eq.~\eqref{chi_r} is deformed into $C_1$,
 $C_2$ (cut contributions) and $C_3$ (pole contribution). For the definition of threshold masses, see Eq.~\eqref{Threshold-mass}. 
} 
\label{Fig-sing}
\end{figure}

The meson screening mass $M_{\xi}^{\rm scr}$ for $\xi$ meson is defined by
\begin{equation}
M_{\xi}^{\rm scr}
=-\lim_{r=|{\bm x}|\rightarrow \infty}\left(\frac{d}{dr}\ln{\zeta_{\xi\xi}(0,{\bm x})}\right), 
\label{scr-mass}
\end{equation}
where the correlation function $\zeta_{\xi\xi}(0,{\bm x})$ 
in coordinate space is 
obtained by the Fourier transformation of the correlation function $\chi_{\xi\xi}(0,{\tilde{q}^2})$ in momentum space as 
\begin{equation}
\zeta_{\xi\xi}(0,{\bm
 x})
=\frac{1}{4\pi^{2}ir}\int^{\infty}_{-\infty}d\tilde{q}\hspace{1ex}\tilde{q}\chi_{\xi\xi}(0,\tilde{q}^2)e^{i\tilde{q}r}; 
\label{chi_r}
\end{equation}
see Fig.~\ref{Fig-sing} to understand the meaning of $\tilde{q}$ integral.

NJL-type effective models have two problems in the calculation of Eq.~\eqref{chi_r}. The first problem stems from the regularization. The 
three-dimensional-momentum cutoff regularization commonly used explicitly
breaks Lorentz invariance, and induces unphysical oscillations in 
$\zeta_{\xi\xi}(0,{\bm x})$~\cite{Florkowski}. This problem can be solved 
by taking the PV regularization~\cite{PV}. We then use the PV 
regularization in this paper. 
The second problem is the fact that direct numerical calculations 
of $\tilde{q}$ integral is quite difficult because the integrand is 
highly oscillating at large $r$ where $M_{\xi}^{\rm scr}$ is
defined. In order to overcome this problem, one can rewrite the
$\tilde{q}$ integral to the complex $\tilde{q}$ integral by using the Cauchy's integral theorem. However,  it is shown in Ref.~\cite{Florkowski} that 
the complex function $\chi_{\xi\xi} (0,\tilde{q}^{2})$ has logarithmic cuts 
in the vicinity of the real $\tilde{q}$ axis. The 
evaluation of the cuts still demands time-consuming numerical calculations. 
Our previous works~\cite{Ishii:2013kaa,Ishii:2015ira} showed 
that the emergence of these logarithmic cuts is avoidable 
by making the ${\bm p}$ integration analytically 
before taking the Matsubara ($n$) summation in 
Eqs. \eqref{chixixi}--\eqref{finte_T_mu}.

Consequently, we obtain the regularized function $I_{3}^{\rm reg}$ as 
an infinite series of analytic functions:  
\begin{eqnarray}
&&I_{3}^{{\rm reg}}(0,\tilde{q}^{2})=iT\sum_{j=1}^{N_c}\sum_{n=-\infty}^\infty\sum_{\alpha=0}^2C_{\alpha}
\nonumber\\
&\times & \int {d^3{\bm p}\over{(2\pi )^3}}
\Bigl[{1\over{{\bm p}^2+\mathcal{M}^2}}
{1\over{({\bm p} + {\bm q})^2+\mathcal{M}^2}}\Bigr]\nonumber\\
&=&{T\over{8\pi \tilde{q}}}\sum_{j,n,\alpha}C_{\alpha}
{\rm Log}\left(
\frac{2\mathcal{M} + i\tilde{q}}
     {2\mathcal{M} - i\tilde{q}}\right)
\label{I_3_v1}
\end{eqnarray}
with a complex valued thermal mass
\begin{equation}
\mathcal{M}(M_\alpha,\omega_n,A_4^{jj},\mu)=\sqrt{M_{\alpha}^2 + (\omega_n + A_{4}^{jj} - i\mu)^2 },
\label{KK_mode}
\end{equation}
where we take the principle value for logarithm in Eq.~\eqref{I_3_v1} 
and the square
root in Eq.~\eqref{KK_mode}. Each term in last line of Eq.~\eqref{I_3_v1} has four cuts starting at
$\tilde{q}=\pm 2i\mathcal{M}(M_\alpha,\omega_n,A_4^{jj},\mu)$ and $\tilde{q}=\pm
2i\mathcal{M}(M_\alpha,\omega_n,-A_4^{jj},-\mu)$, as shown in
Fig.~\ref{Fig-sing}. For later convenience, we define the threshold mass 
$M_{\rm th}$
and the decay width $\Gamma_{\rm th}$ by the $\mathcal{M}$ located at the 
lowest branch point in the 
upper-half plane: Namely  
\begin{equation}
2\mathcal{M}_{\rm lowest} \equiv  M_{\rm th} - i\frac{\Gamma_{\rm th}}{2},
\label{Threshold-mass}
\end{equation} 
where $M_{\rm th}$ ($\Gamma_{\rm th}$) is the real (imaginary) part of
$2 \mathcal{M}_{\rm lowest}$. 
Meson screening mass $M_{\xi}^{\rm scr}$ is a pole of
$\chi_{\xi\xi}$ and is calculated by 
\begin{equation}
\left.\left[ 1-2G_{\rm S}(\Phi)\Pi_{\xi}(0,\tilde{q}^2)\right] 
\right|_{\tilde{q}=iM_{\xi}^{\rm scr}} = 0, 
\label{scr-eq}
\end{equation}
when the pole is located below the lowest branch point. This
condition leads to~\cite{Ishii:2013kaa} 
\begin{equation}
M_{\xi}^{\rm scr}\le M_{\rm th}.
\label{threshold mass}
\end{equation}

\section{Numerical Results}
\label{Numerical Results}

\subsection{Deconfinement and chiral transition lines in $\theta$-$T$ plane}
\label{Chiral and Polyakov-loop}

\begin{figure}[htbp]
\begin{center}
  \includegraphics[width=0.45\textwidth]{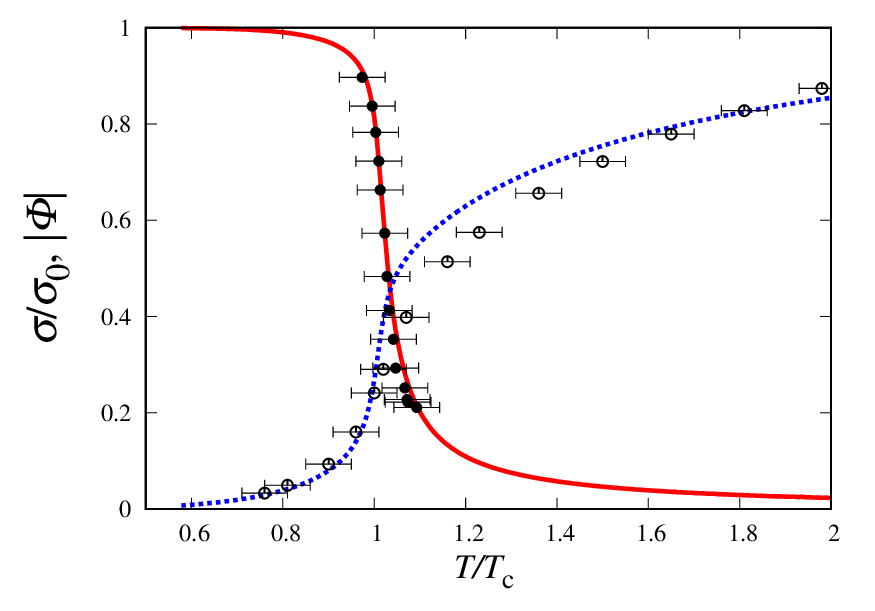}
\end{center}
\caption{$T$ dependence of the chiral condensate $\sigma$ and 
the absolute value $|\Phi|$ of Polyakov loop for $\theta=0$. 
The horizontal axis is scaled by the  mean value $T_{\rm c}=173$ MeV of LQCD results on the chiral transition temperature
at $\theta=0$~\cite{Chiral-condensate-Karsch}. 
The $\sigma$ is normalized by the value ($\sigma_0$) at $T=0$. 
LQCD data  are taken 
from Refs.~\cite{Polyakov-loop-Kaczmarek,Tc-Karsch}. 
Note that the 10 \% errors come from those of $T_{\rm c}$. 
}
\label{Fig-sigma-Phi}
\end{figure}

Figure \ref{Fig-sigma-Phi} shows $T$ dependence of $\sigma$ and $|\Phi|$ 
for the case of $\theta=0$. 
The EPNJL-model results with the parameter set of Table~\ref{Model parameters}
well simulate LQCD data~\cite{Polyakov-loop-Kaczmarek,Tc-Karsch}. 
This means that the present EPNJL model is reliable at least for 
$\theta=0$. 

\begin{figure}[htbp]
\begin{center}
  \includegraphics[width=0.45\textwidth]{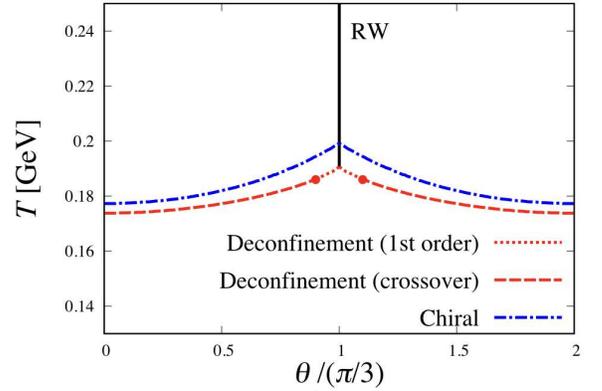}
\end{center}
\caption{
Deconfinement and chiral transition lines in the imaginary-$\mu$ region. The dashed line (dot-dash line) stands
 for the deconfinement (chiral) transition line. The RW transition
 line is denoted by the solid line. At the point $(\theta,T)=(\pi/3\pm 0.084,187~[{\rm  MeV}])$, the
 deconfinement transition becomes the second order from the first order. 
The locations are shown by two dots.
}
\label{Transition-line}
\end{figure}

Figure~\ref{Transition-line} shows the deconfinement and chiral
transition lines in the imaginary-$\mu$ region, where the transition
temperatures are determined from peak positions of chiral and
Polyakov-loop susceptibilities. 
$\theta$ dependence of the transition lines are well fitted 
in $0\le\theta\le\pi/3$ by using 
\beq
\frac{T_{\rm c}^{\rm X}(\theta)}{T_{\rm c}^{\rm X}} 
= 1 + c_1^{\rm X}\theta^2 + c_2^{\rm X}\theta^4, 
\label{Tc_approx}
\eeq
where the  superscript ``${\rm X}={\rm d}$'' means the deconfinement transition 
and ``${\rm X}=\chi$'' corresponds to the chiral transition. 
The results of the fitting are summarized in Table.~\ref{Tc_parameter-0}.

\begin{table}[h]
\begin{center}
\caption
{Parameter sets for the deconfinement- and 
chiral-transition lines.
}
\begin{tabular}{lccccc}
\hline\hline
&$T_{\rm c}^{\rm X}(0)$~[MeV]
&$c_1^{\rm X}$
&$c_2^{\rm X}$
\\
\hline
Deconfinement
&174
&0.064
&0.019
\\
Chiral
&177
&0.090
&0.020
\\
\hline
\end{tabular}
 \label{Tc_parameter-0}
\end{center}
\end{table}

\subsection{$\theta$ dependence of $\pi$ and $\sigma$ meson screening masses}

First we have confirmed that $\pi$- and $\sigma$-meson screening masses have 
the RW periodicity and charge symmetry:
\begin{equation}
 M_{\xi}^{\rm scr}\left(\theta\right) =  M_{\xi}^{\rm
  scr}\left(\theta + \frac{2\pi k}{3}\right),~
 M_{\xi}^{\rm scr}\left(\theta\right) =  M_{\xi}^{\rm
  scr}\left(-\theta\right)
\label{scr-RW-CC-sym}
\end{equation}
for $\xi=\pi,\sigma$, where $k$ is an arbitrary integer. 
This result stems from the fact that
Eqs. \eqref{I1}--\eqref{finte_T_mu} and the threshold mass $M_{\rm th}$ 
are invariant under the extended $Z_3$ transformation 
defined by
Eq.~\eqref{extended-Z3}.



In the next subsection, we will extrapolate the meson screening masses 
from $\mu=i\mu_{\rm I}$ to $\mu=\mu_{\rm R}$. 
For this purpose, we first fit our model results with the polynomial 
function,  
\begin{equation}
\frac{M_{\xi}^{\rm scr}(T,i\mu_{\rm I})}{T} 
= \sum_{n=0}^{n_{\rm max}}a^{(n)}_\xi(T)\theta^{2n},
\label{fit-func}
\end{equation}
in $0\le\theta\le\pi/3$. We take $n_{\rm max}=1,2,3,4$ in order to 
confirm convergence of the expansion.  
$\theta$ dependence of $M_{\pi}^{\rm scr}$ and $M_{\sigma}^{\rm scr}$ 
is well fitted with $n_{\rm max}=4$. 
In this procedure, $\theta$
is varied in the range $0\le\theta\le \pi/3$, although 
$T$ is fixed.

We consider the following two cases: 
 
\begin{itemize}
\item[(A)]
$T=250$ MeV in Fig.~\ref{Transition-line}: 
The system is in both the deconfinement and the chiral-symmetry 
restored phase for any $\theta$, since $T \ge T_{\rm c}^{\chi}(\pi/3)$. 
 
\item[(B)]
$T=180$ MeV in Fig.~\ref{Transition-line}:  This case satisfies 
$T_{\rm c}^{\chi}(0)\le T\le T_{\rm RW}$.
The system is in the deconfinement phase for $0\le \theta\le 0.697$ but
          in the confinement phase in $0.697\le \theta\le \pi/3$. The
          system is in the chiral-symmetry restored phase for $0\le
          \theta\le 0.403$ but in the chiral-symmetry broken phase 
for $0.403\le \theta\le \pi/3$.  
\end{itemize}

\begin{figure}[htbp]
\begin{center}
  \includegraphics[width=0.5\textwidth]{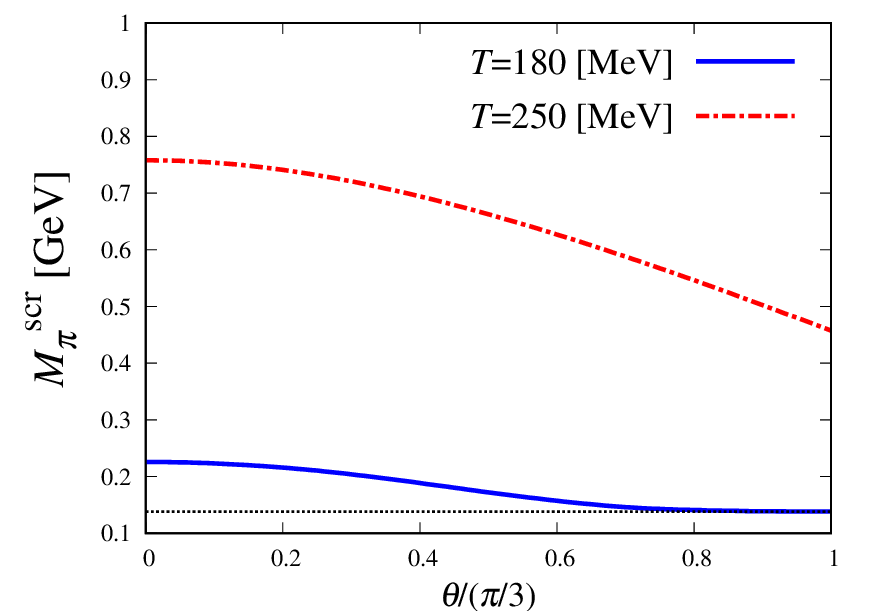}
\end{center}
\caption{$\theta$ dependence of $M_{\pi}^{\rm scr}$ for two
 cases (A) and (B). The dotted line denotes $\pi$ meson screening 
masses at vacuum.}
\label{pion-scr}
\end{figure}

Figure~\ref{pion-scr} explains 
$\theta$ dependence of $\pi$-meson screening 
masses for two cases (A) and (B). The $M_{\pi}^{\rm scr}$ 
monotonically decrease as $\theta$ increases for two cases (A) and (B).

\begin{figure}[htbp]
\begin{center}
  \includegraphics[width=0.5\textwidth]{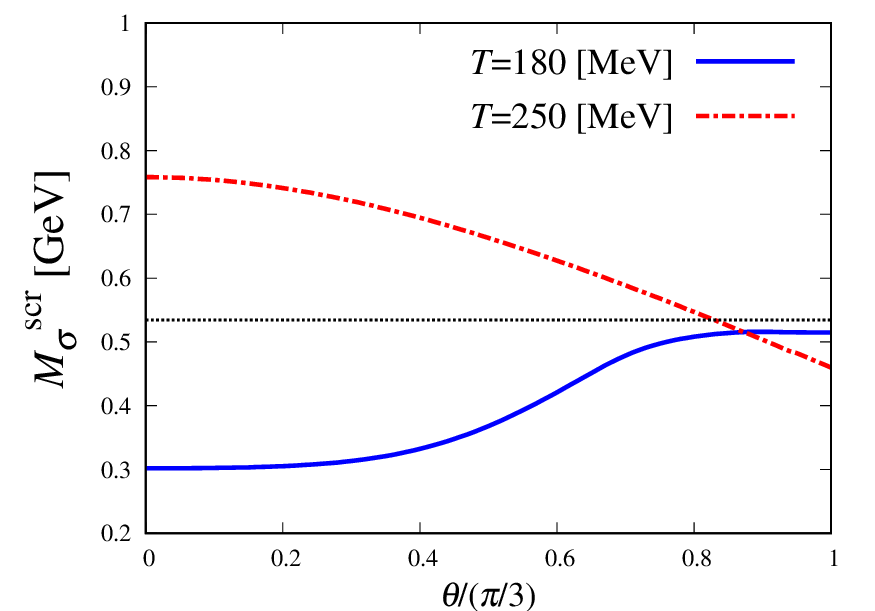}
\end{center}
\caption{$\theta$ dependence of $M_{\sigma}^{\rm scr}$ for two 
 cases (A) and (B). The dotted line denotes $\sigma$ meson screening 
masses at vacuum.
}
\label{sigma-scr}
\end{figure}

Figure~\ref{sigma-scr} shows $\theta$ dependence of $\sigma$-meson screening 
masses for two cases (A) and (B).
The $M_{\sigma}^{\rm scr}$ have  
non-monotonic $\theta$ dependence for case (B). As for case (A), the $\pi$- and $\sigma$-meson screening 
masses agree with each other due to the chiral symmetry restoration.

\subsection{Extrapolation from $\mu_{\rm I}$ to $\mu_{\rm R}$ region}

We compare the extrapolating result with the direct one  
for finite $\mu_{\rm R}$ in order to confirm applicability of the analytic 
continuation. One can easily make the analytic
continuation by replacing $\theta$ with 
$-i \mu_{\rm R}/T$:
\begin{eqnarray}
\frac{M_{\xi}^{\rm scr}(T,\mu_{\rm R})}{T} 
= \sum_{n=0}^{n_{\rm max}}(-1)^{n}a^{(n)}_\xi(T)
\left(\frac{\mu_{\rm R}}{T}\right)^{2n}.
\label{extrapolnation-series}
\end{eqnarray}

\begin{figure}
\begin{center}
  \includegraphics[width=0.5\textwidth]{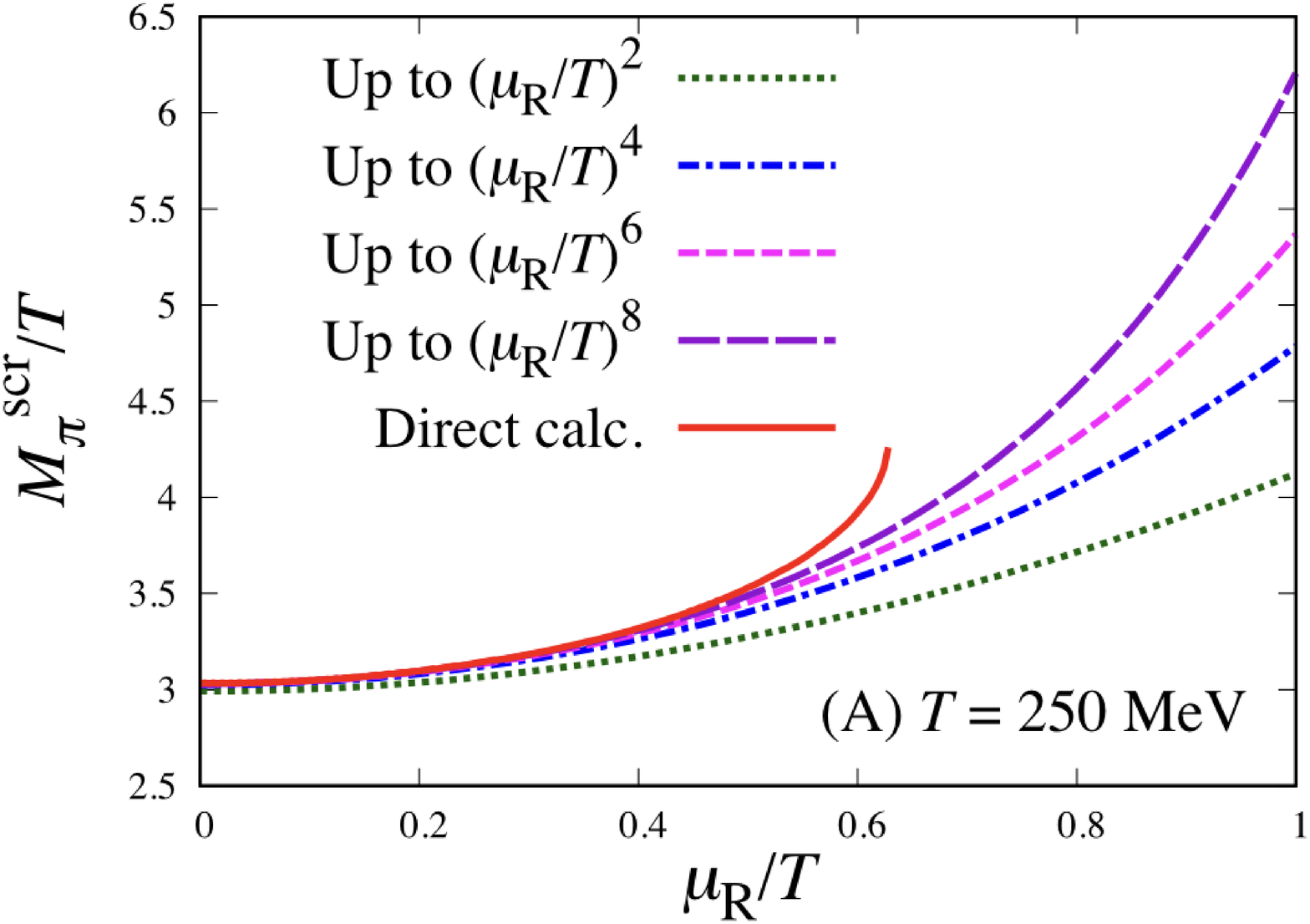}
\end{center}
\begin{center}
  \includegraphics[width=0.5\textwidth]{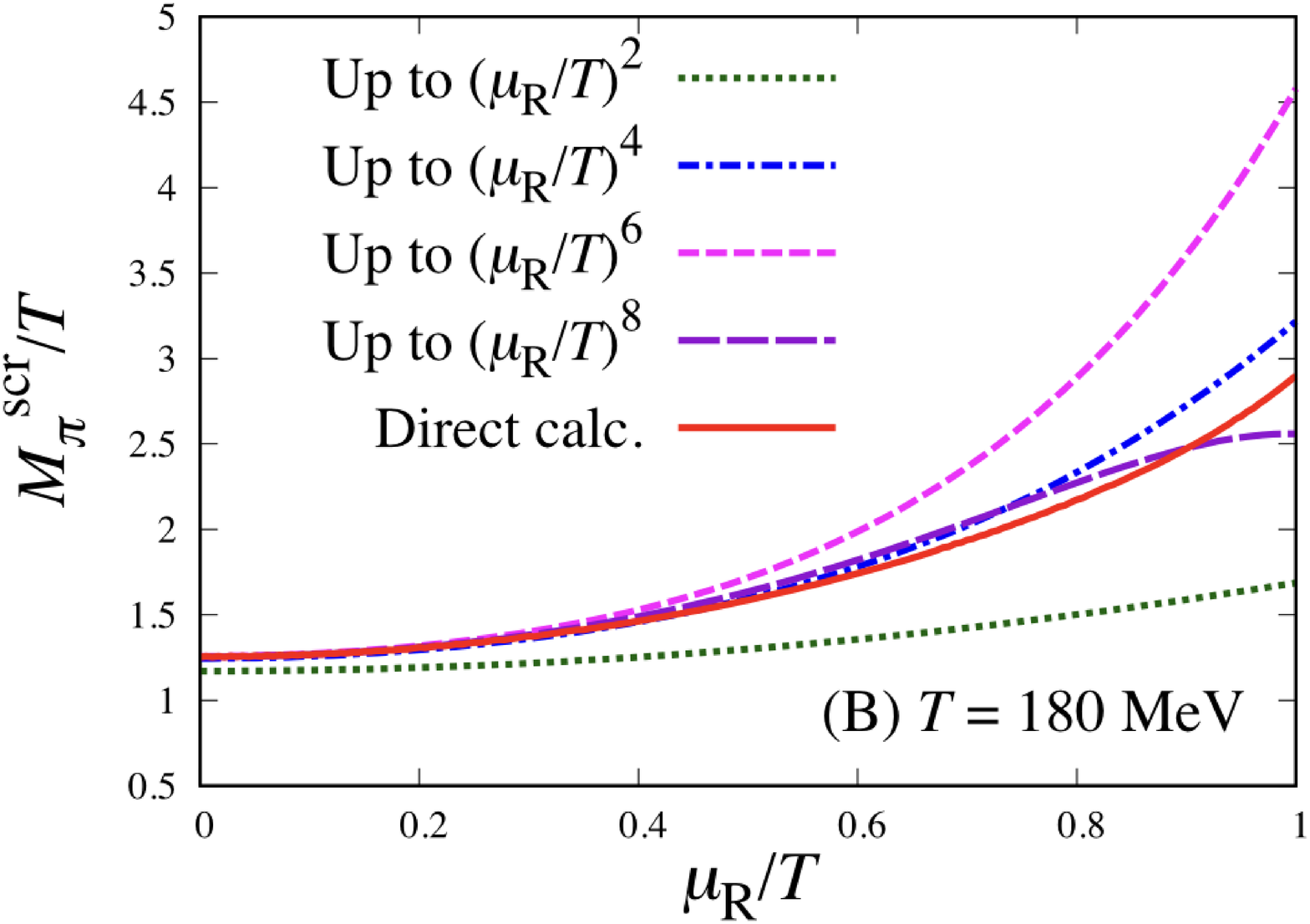}
\end{center}
\caption{Comparison between the extrapolating and the direct results
 on $\mu_{\rm R}/T$ dependence of $M_\pi^{\rm scr}$. We draw direct-result lines only when the inequality 
$M_{\pi}^{\rm scr}<M_{\rm th}$ in Eq.~\eqref{threshold mass} is satisfied. 
}
\label{Ext-pi-T180-250-range1.047-real}
\end{figure}

Figure~\ref{Ext-pi-T180-250-range1.047-real} explains 
$\mu_{\rm R}$ dependence of $\pi$-meson screening 
masses for two cases (A) and (B). 
In $\mu_{\rm R}/T \lsim 0.4$, the $M_{\pi}^{\rm scr}$ converge to the
direct results as $n_{\rm max}$ increases for both the two cases.

\begin{figure}[htbp]
\begin{center}
  \includegraphics[width=0.5\textwidth]{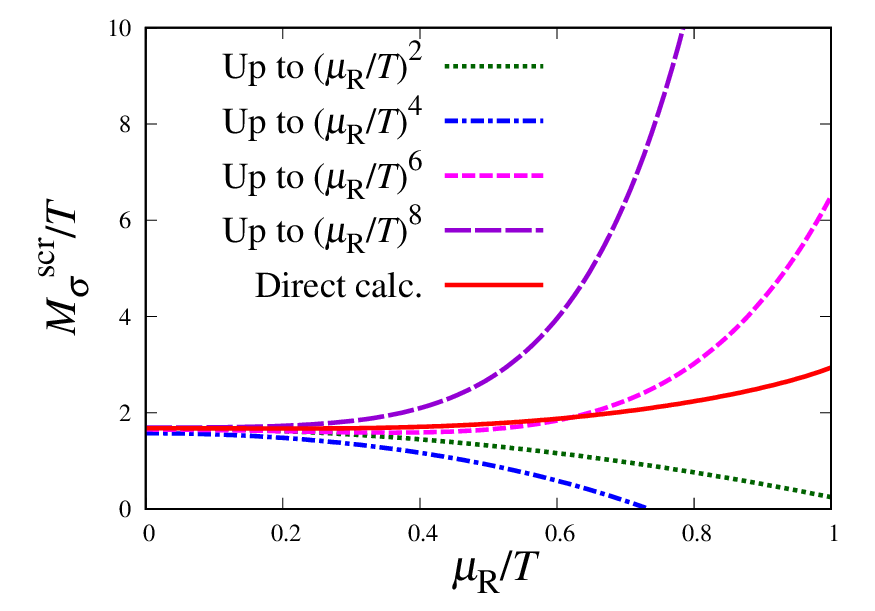}
\end{center}
\caption{
Comparison between the extrapolating and the direct results
 on $\mu_{\rm R}/T$ dependence of $M_\sigma^{\rm scr}$ in case (B), i.e., $T=180$ MeV.
}

\label{Ext-sigma-T180-range1.047}
\end{figure}

Figure~\ref{Ext-sigma-T180-range1.047} shows 
$\mu_{\rm R}/T$ dependence of $M_\sigma^{\rm scr}$ for case (B), i.e., $T=180$ MeV. We skip case (A) since chiral symmetry is restored in case (A), and $\theta$ dependence of
$M_\sigma^{\rm scr}$ is almost same as that of $M_\pi^{\rm scr}$. 
The extrapolating results tends to the direct ones for $0\le
\mu_{\rm R}/T\lsim 0.4$, and the deviation in $0.4\le
\mu_{\rm R}/T$ can not be improved by
taking the higher order terms.

The origin of the deviation can be understood when one considers the
relation between $\sigma$-meson screening mass and chiral
susceptibility. Equation~\eqref{scr-mass} indicates that 
the inverse of $M_\sigma^{\rm scr}$ corresponds to the correlation
length in the fluctuation of $\langle\bar{\psi}(x)\psi(x)\rangle$; 
see Ref.~\cite{Fujii:2003bz} for the further explanation, and note that
screening mass is referred to be the frequency of ``sound mode'' there.   
Hence $M_{\sigma}^{\rm scr}$ is related to 
the chiral susceptibility $\chi_\sigma$ as 
\bea
M_\sigma^{\rm scr}\propto\chi_\sigma^{-1/2}. 
\label{Eq:sigma-sus}
\eea
Particularly for the chiral limit, $\mu_{\rm R}$ and $\mu_{\rm I}$
dependence of $M_\sigma^{\rm scr}$ is non-analytic 
on the chiral phase transition line in ${\mu}_{\rm R}$--$T$ and ${\mu}_{\rm I}$--$T$ plane, 
since $\chi_\sigma$ is non-analytic on the chiral phase transition
line. As for finite quark mass, a remnant of the non-analycity makes the accuracy of 
the analytic continuation less accurate. 

\subsection{Phase-transition-line extrapolation}

\begin{figure}[htbp]
\begin{center}
  \includegraphics[width=0.5\textwidth]{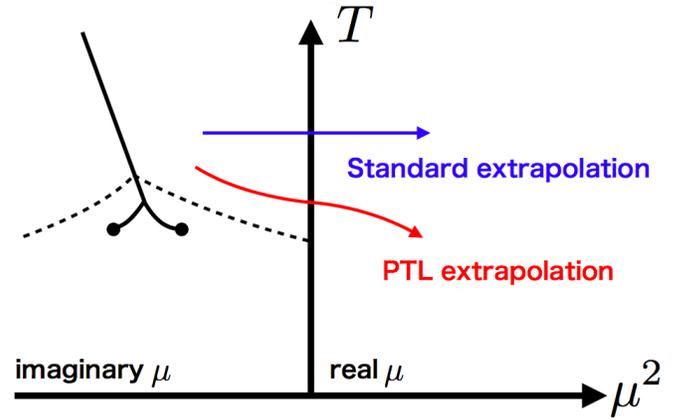}
\end{center}
\caption{A schematic figure of PTL extrapolation and standard-extrapolation. 
The arrows stand  for the  standard extrapolation and the PTL
 extrapolation. Transition line for chiral symmetry restoration is denoted
 by the dotted line.
}
\label{PTL}
\end{figure}

We propose the new extrapolation method by modifying a trajectory
of $(T,\theta)$ in fitting. In standard extrapolation, $\theta$ is
varied with fixed $T$. In new method, we also vary $T$ so that the trajectory
runs along the phase transition line. We then assume $\theta$ dependence
of $T$ as   
\begin{equation}
T=T_{\rm PTL}^{\rm X}(\theta) = R\times T_{\rm c}^{\rm X}(\theta) 
\label{T_PTL}
\end{equation}
with any constant $R$ that is introduced to cover the $\theta$-$T$
plane; see Fig.~\ref{PTL} for the understanding. The symbol ${\rm X}$ means
the chiral transition (${\rm X}=\chi$) or deconfinement transition
(${\rm X}={\rm d}$). 
In this paper, we refer to the modified extrapolation as
``phase-transition-line (PTL) extrapolation''. 

From now on, we consider the chiral transition ($X=\chi$). 
We fit $\theta$ dependence of $\sigma$-meson
screening masses with a polynomial series:
\begin{equation}
\frac{M_{\sigma}^{\rm scr}(\theta)}
{T_{\rm PTL}^{\chi}(\theta)} 
= \sum_{n=0}^{n_{\rm max}}b^{(n)}_\sigma(R)\theta^{2n}.
\label{fit-func-PTL}
\end{equation}
In Eq.~\eqref{T_PTL}, the
extrapolation line does not pass
through the chiral transition line, we can use all range of $\theta$
for fitting, i.e.,
$0\le\theta\le\pi/3$.

\begin{figure}[htbp]
\begin{center}
  \includegraphics[width=0.5\textwidth]{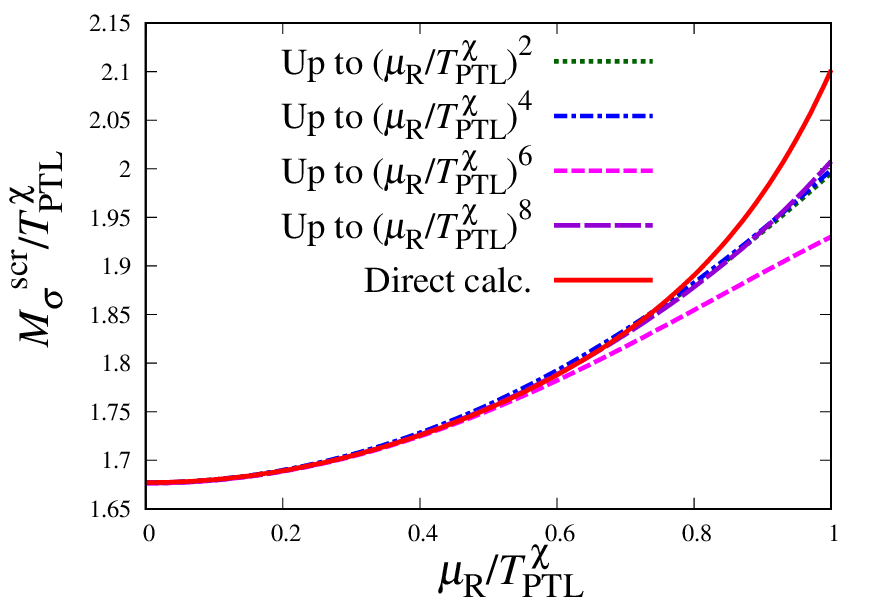}
\end{center}
\caption{$\mu_{\rm R}/T_{\rm PTL}^{\chi}$ dependence of $\sigma$-meson screening mass for
 $T_{\rm PTL}^{\chi}(0)=180$ MeV.}
\label{Ext-sigma-range-TL}
\end{figure}

We then extrapolate $M_\sigma^{\rm scr}(\theta)$ and $T_{\rm
PTL}^{\chi}(\theta)$ from finite $\mu_{\rm I}$ region to $\mu_{\rm R}$ region. Figure~\ref{Ext-sigma-range-TL} shows the
comparison between the direct result and the extrapolating ones for 
$M_{\sigma}^{\rm scr}$, where we set $T_{\rm PTL}^{\chi}(0)=180$ MeV. The extrapolating results rapidly converge to direct-calculated one 
in $\mu_{\rm R}/T_{\rm PTL}^{\chi} \lsim 0.8$. 
The PTL extrapolation yields better agreement than 
the standard extrapolation. 

We also check the reliability of extrapolation by
estimating the radius of convergence in
Eq.~\eqref{fit-func-PTL} based on the d'Alembert ratio test. The
coefficients $b_\sigma^{(n)}$ in Eq.~\eqref{fit-func-PTL} are summarized in Table.~\ref{Coefficient}. The radius of
convergence $r_\sigma$ is calculated by $r_\sigma\equiv\sqrt{b^{(n_{\rm
max}-1)}_\sigma/b^{(n_{\rm max})}_\sigma}\simeq 0.84$, whose value is
consistent with the upper bound of the agreement region.

\begin{table}[htbp]
\begin{center}
\caption
{
Coefficients and convergence radii for $M_{\sigma}^{\rm scr}$ 
and $M_{\pi}^{\rm scr}$ with $n_{\rm max}=4$.}
\begin{tabular}{lccccccc}
\hline\hline
&
&$b_{\xi}^{(0)}$
&$b_{\xi}^{(1)}$
&$b_{\xi}^{(2)}$
&$b_{\xi}^{(3)}$
&$b_{\xi}^{(4)}$
&$r_\xi$
\\
\hline
&$\sigma$ meson
&$1.677$
&$-0.312$
&$-0.010$
&$-0.012$
&$0.017$
&$0.84$
\\
&$\pi$ meson 
&$1.254$
&$-0.327$
&$0.168$
&$-0.090$
&$0.0184$
&$2.21$
\\
\hline
\end{tabular}
 \label{Coefficient}
\end{center}
\end{table}

\begin{figure}[htbp]
\begin{center}
  \includegraphics[width=0.5\textwidth]{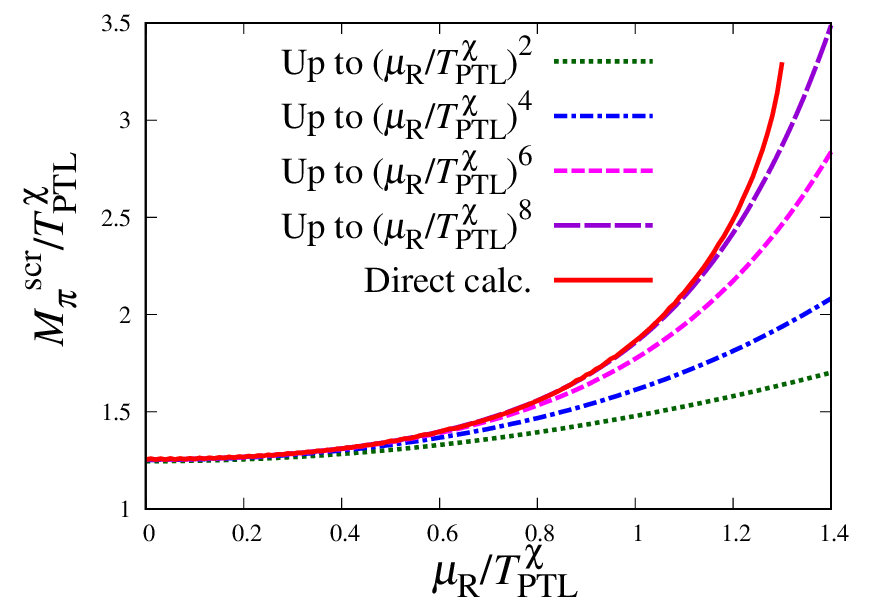}
\end{center}
\caption{$\mu_{\rm R}/T_{\rm PTL}^{\chi}$ dependence of $\pi$-meson screening mass for $T_{\rm PTL}^{\chi}(0)=180$ MeV.}
\label{Ext-pi-range-TL}
\end{figure}

Parallel discussion is possible for $M_{\pi}^{\rm scr}$, as shown in 
Fig. \ref{Ext-pi-range-TL}. We can obtain good agreement between direct results and
extrapolating ones for $\mu_{\rm R}/T_{\rm PTL}^{\chi}\le \pi/3$.

\section{Summary}
\label{Summary}

We first showed a method of calculating 
screening masses for finite $\mu_{\rm R}$ and $\mu_{\rm I}$ 
in the framework of the 2-flavor EPNJL model. 

Next, we investigated how reliable 
the imaginary-$\mu$ approach is 
for $M_{\pi}^{\rm scr}$ and $M_{\sigma}^{\rm scr}$ 
by comparing ``the results extrapolated from imaginary $\mu$'' with
``those  calculated directly in real $\mu$''. 
In the standard extrapolation, the agreement between the direct and the
extrapolating results is seen in $\mu_{\rm R}/T \lsim 0.4$ for $M_{\pi}^{\rm
scr}$ and 
 $M_{\sigma}^{\rm scr}$ for $T=180$~and $250$ MeV. Especially for
 $\sigma$ meson, the disagreement in $0.4\lsim
\mu_{\rm R}/T$ can not be improved by
taking higher order terms.

We can understand the difficulty of
extrapolation when one remembers that $M_\sigma^{\rm scr}$ is nothing
but the inverse of correlation length in fluctuation of local chiral
condensate. The $M_\sigma^{\rm scr}$ is thus related with the chiral
susceptibility $\chi_\sigma$ as $M_\sigma^{\rm
scr}\propto\chi_\sigma^{-1/2}$. When one set quark mass to zero,
$\chi_\sigma$ becomes non-analytic on the chiral transition line
$T=T_{\rm c}^{\chi}(\theta)$, and so does $M_\sigma^{\rm scr}$. Even for finite quark mass, a remnant of this non-analicity 
makes the accuracy of extrapolation
less accurate, since quark mass is much smaller to temperature 
and negligible
around the chiral phase transition. 
 This indicates that the simple
extrapolation is not useful for $M_\sigma^{\rm
scr}(T,\mu_{\rm R})$.

In order to circumvent this problem, we propose the PTL extrapolation. In the method, the agreement between the direct and the
extrapolating results is seen in $\mu_{\rm R}/T_{\rm PTL}^{\chi} \lsim 0.8$ for
$M_{\sigma}^{\rm scr}$ and in $\mu_{\rm R}/T_{\rm PTL}^{\chi} \lsim \pi/3$ for 
 $M_{\pi}^{\rm scr}$ with $T_{\rm PTL}^{\chi}(0)=180$~MeV. The
 extrapolating results tend to the direct results as
 higher order terms are taken into account. 
The PTL extrapolation thus 
makes better extrapolating results than the standard one.

The difficulty of the simple extrapolation may be in common with other scalar,
vector and pseudovector mesons composed of u and d quarks, since these meson masses are sensitive
to the chiral transition. The application of PTL extrapolation to such mesons
is thus interesting as a future perspective.

\noindent
\begin{acknowledgments}
The authors thank to Kouji Kashiwa and Junpei Sugano for fruitful discussion. M. I., H. K.,
 and M. Y. are supported by Grants-in-Aid for Scientific Research
 (No.~27-3944, No.~17K05446 and No.~26400278) from the Japan Society for
 the Promotion of Science (JSPS). 
\end{acknowledgments}

\noindent
\appendix*
\section{The relation between $A_4$ and $\Phi,\bar{\Phi}$}
\label{A4-from-Polyakov-loop}

The diagonal components $A_4^{11},A_4^{22},A_4^{33}$ of the gluon field are 
related with the Polyakov loop $\Phi$ and its conjugate $\bar{\Phi}$ as 
\begin{eqnarray}
\Phi 
&=& \frac{1}{3}(\phi_1 + \phi_2 + \phi_3),
\\ 
\bar{\Phi} 
&=& \frac{1}{3}(\phi_1^{*} + \phi_2^{*} + \phi_3^{*})
= \frac{1}{3}(\phi_1\phi_2 + \phi_2\phi_3 + \phi_3\phi_1)
\nonumber\\
\label{Phi-A4}
\end{eqnarray}
with $\phi_j \equiv \exp{(iA_4^{jj}/T)}$ ($j=1,2,3$). Furthermore, the traceless condition for $A_4$ leads to 
\begin{equation}
\phi_1\phi_2\phi_3 = 1.  
\label{traceless}
\end{equation}
One can confirm that $\phi_1,\phi_1,\phi_3$ are solutions of
following eqution: 
\begin{equation}
\phi^3 - 3\Phi \phi^2 + 3\bar{\Phi}\phi - 1 = 0.
\end{equation}
by considering Vieta's formulas. Once we get $\Phi$ and $\bar{\Phi}$, we can obtain $\phi_1,\phi_2,\phi_3$
by solving above equation analytically and get the gluon field as
$A_4^{jj} = -iT\log{\phi_{j}}$. The relation between
$A_4^{jj}$ and $ \phi_j$ has an
ambiguity coming from the replacement $A_4\to A_4 + 2n\pi T$ for integer
$n$, but this ambiguity does not change any physical observables and
we simply assume $n=0$. If we take the approximation $\Phi\simeq
\bar{\Phi}$, we can simply obtain the gluon fields as 
\begin{equation}
A_4^{11}=-A_{4}^{22}=T\cos^{-1}{\left(\frac{3\Phi-1}{2}\right)},~
A_4^{33}=0. 
\end{equation}


\end{document}